\documentclass[runningheads]{llncs}

\usepackage{url}
\usepackage{hyperref}
\usepackage{tabularx}
\usepackage{multirow}
\usepackage{color}
\usepackage{booktabs}
\usepackage[square,comma,numbers,sort&compress,sectionbib]{natbib}
\usepackage{enumerate}
\usepackage{diagbox}

\usepackage{graphicx}

\usepackage{fancyhdr}

\fancypagestyle{specialfooter}{%
  \fancyhf{}
  
  \fancyfoot[C]{\vspace{1cm} \hspace*{-.2\textwidth}\parbox{1.4\textwidth}{\small This is the author's version of the work. It is posted here for your personal use. The definitive version is published in: \\ \emph{Proceedings of the 46th European Conference on Information Retrieval} (ECIR '24), March 24--28, 2024, Glasgow, Scotland}}
}

\begin{document}
\mainmatter

\title{Towards Reliable and Factual Response Generation: Detecting Unanswerable Questions in Information-Seeking Conversations}

\titlerunning{Towards Reliable and Factual Response Generation}  
\author{Weronika Łajewska\orcidID{0000-0003-2765-2394} \and \mbox{Krisztian Balog}\orcidID{0000-0003-2762-721X}}

\authorrunning{Łajewska and Balog}

\tocauthor{Weronika Łajewska, Krisztian Balog}

\institute{University of Stavanger, Stavanger, Norway,\\
\email{\{weronika.lajewska, krisztian.balog\}@uis.no}}

\maketitle

\begin{abstract}
Generative AI models face the challenge of hallucinations that can undermine users' trust in such systems. We approach the problem of conversational information seeking as a two-step process, where relevant passages in a corpus are identified first and then summarized into a final system response. This way we can automatically assess if the answer to the user's question is present in the corpus. Specifically, our proposed method employs a sentence-level classifier to detect if the answer is present, then aggregates these predictions on the passage level, and eventually across the top-ranked passages to arrive at a final answerability estimate. For training and evaluation, we develop a dataset based on the TREC CAsT benchmark that includes answerability labels on the sentence, passage, and ranking levels. We demonstrate that our proposed method represents a strong baseline and outperforms a state-of-the-art LLM on the answerability prediction task.
 
\end{abstract}

\thispagestyle{specialfooter}

\section{Introduction}

Conversational information seeking (CIS) systems allow users to fulfill their complex information needs via a sequence of interactions. This problem is often approached as a passage retrieval task~\citep{Luan:2021:Transactionsa, Dalton:2019:TRECa}, rather than employing generative AI techniques, to allow for the grounding of responses in supporting documents and to avoid hallucinations. However, the ultimate goal is to return informative, concise, and reliable answers instead of top-ranked passages. In an ideal scenario, when the passages from the top of the ranking answer the question, the task of response generation boils down to summarization~\citep{Owoicho:2022:TRECa}. However, it is often the case that the answer to the user's question is not contained in the top retrieved passages. In such cases, summaries generated from those passages would result in hallucinations~\citep{Tang:2023:arXiva, Cao:2016:COLINGa}. 

In this paper, we make the first step towards reliable and factual conversational response generation. We propose a mechanism for detecting unanswerable questions for which the correct answer is not present in the corpus or could not be retrieved. More specifically, given a set of top-ranked passages that have been identified as most relevant to the given question, we predict if the question can be answered (at least partially) based on information contained in those passages. This enables us to move beyond the notion of passage relevance and focus more on the actual presence of the information that answers the question. Introducting this additional step of answerability prediction in the CIS pipeline, to be performed after the passage retrieval and before the response generation steps, could help mitigate hallucinations and factual errors. 
It would enable the system to transparently communicate to the user if the answer to the query could not be found, instead of generating a response from only marginally relevant passages.

Unanswerability detection has been addressed in the context of machine reading comprehension~\citep{Huang:2019:CoNLLa, Hu:2018:arXiva} and question answering~\citep{Sulem:2022:NAACL-HLT, Choi:2018:EMNLPa, Rajpurkar:2018:ACLa, Reddy:2019:Trans.a}, both of which differ significantly from the conversational search setup. Information-seeking dialogues pose additional challenges, such as open-ended questions requiring descriptive answers, indirect answers requiring an inference or background/context knowledge, and complex queries with partial answers spread across passages. Therefore, unanswerability detection is a novel, still unsolved task in CIS, and, to the best of our knowledge, no public dataset exists for this problem.  

As our first main contribution, we develop a dataset, based on the TREC CAsT benchmark, to train and evaluate methods for question answerability prediction. Utilizing an existing resource of snippet-level answer annotations~\citep{Lajewska:2023:CIKM}, our dataset provides answerability labels on three levels: (1) sentences, (2) passages, and (3) rankings (i.e., top-ranked passages retrieved by a CIS system). 
Notably, we generate input passage rankings with various degrees of difficulty in answerability prediction, mixing passages that contain answers with those with no answers, in a controlled way. As a result, passage rankings range from all passages containing an answer to ``no answer found in the corpus.'' 

As our second main contribution, we provide a baseline approach for predicting answerability based on an input ranking.  
Our proposed approach predicts which sentences from the top-ranked passages contribute to the answer and aggregates the obtained answerability scores on the passage and ranking levels. We demonstrate that this simple method provides a strong baseline that outperforms ChatGPT-3.5 on the same task. 
Further, we show that augmenting our dataset with additional training samples for unanswerable question detection (from the SQuAD 2.0 dataset~\citep{Rajpurkar:2018:ACLa}) does not improve ranking-level answerability prediction in conversational search, underscoring the distinct character of this task.
Our benchmark dataset (CAsT-answerability) as well as the implementation of our proposed method are made publicly available at \url{https://github.com/iai-group/answerability-prediction}.

\section{Related Work}
\label{sec:related}

Research on information-seeking conversations is largely driven by test collections developed as part of the TREC Conversational Assistance Track (CAsT)~\citep{Dalton:2019:TRECa, Dalton:2020:TRECa, Dalton:2021:TRECa, Owoicho:2022:TRECa}. Unlike generative AI approaches, answers in this benchmark are grounded in passages, hence the problem boils down to that of conversational passage retrieval~\citep{Pradeep:2021:arXiv,Vakulenko:2021:WSDMa, Luan:2021:Transactionsa}. Aggregating results from top-ranked passages into a single answer is an open problem~\citep{Bolotova:2023:ACL} that has been first piloted in the 2022 edition, where a subtask of generating summaries from retrieved results was introduced~\citep{Owoicho:2022:TRECa}. 
Ren et al.~\cite{Ren:2021:TOISa} propose an approach for response generation divided into three stages: (optional) query rewriting, finding supporting sentences in results displayed on a SERP, and summarizing them into a short conversational response. While the authors acknowledge the problem of unanswerability in conversational search, they do not address it in their proposed approach.
In this paper, we aim to fill that gap. 

The problem of unanswerability has been addressed in the context of machine reading comprehension (MRC)~\citep{Huang:2019:CoNLLa, Hu:2018:arXiva} and extractive question-answering (QA)~\citep{Asai:2021:ACL-IJNLPa, Liao:2022:SIGIRa, Godin:2019:NAACL-HLTa}. 
Solutions proposed include 
answerability prediction using prompt-tuning~\citep{Liao:2022:SIGIRa}, 
modeling high-level semantic relationships between objects from question and context~\citep{Huang:2019:CoNLLa},
and combining the output of reading and verification modules in MRC systems~\citep{Hu:2018:arXiva,Zhang:2020:arXivb}.
Our proposed solution is based on a sentence-level classifier that is learned on CIS-specific training data, and can further be augmented with QA answerability data.

\section{Dataset}
\label{sec:expsetup}

\begin{table}[t]
    \small
    \caption{Statistics for the CAsT-answerability dataset.}
    \label{tab:dataset_stats}
    \centering
    \begin{tabular}{l|r|r}
    \multicolumn{1}{c|}{} & \multicolumn{2}{c}{\textbf{Answerable?}} \\
    \cline{2-3}
    \multicolumn{1}{c|}{} & \textbf{Yes} & \textbf{No} \\
        \hline
        \#question-sentence pairs (train+test) & 6,395 & 19,043 \\
        \#question-passage pairs (train+test) & 1,778 & 1,932 \\
        \#question-ranking pairs (test) & 4,035 & 504 \\
    \end{tabular}
\end{table}

This paper builds upon the CAsT-snippets dataset~\citep{Lajewska:2023:CIKM},\footnote{\url{https://github.com/iai-group/CAsT-snippets}}
which extends the TREC CAsT'20 and '22 datasets with snippet-level annotations for the top-retrieved results.
Specifically, it contains annotations of information nuggets defined as ``minimal, atomic units of relevant information''~\citep{Pavlu:2012:WSDMa}, representing key pieces of information required to answer the given question. 
Snippets in the dataset are identified for every question in the 5 most relevant passages according to ground truth judgments. 
To balance the collection, we also include 5 randomly selected non-relevant passages to each question.
The resulting dataset, named \emph{CAsT-answerability}, contains around 1.8k answerable and 1.9k unanswerable question-passage pairs. 
We further consider answerability on the level of sentences and on the level of rankings, as follows.
For sentence-level answerability, we leverage annotations of information nuggets from the CAsT-snippets dataset as follows:
each sentence that overlaps with an information nugget, as per annotations in the originating CAsT-snippets dataset, is labeled as 1 (answer in the sentence), otherwise as 0 (no answer in the sentence).

For ranking-level answerability, which is the ultimate task we are addressing, we consider different input rankings, i.e., sets of $n=3$ passages, for the same input question. 
Specifically, for each unique input test question (38), we generate all possible $n$-element subsets of passages available for this question (both containing and not containing an answer), thereby simulating passage rankings of varying quality.
These rankings represent inputs with various degrees of difficulty for the same question, ranging from all passages containing an answer to a single passage with an answer to ``no answer found in the corpus.''
This yields a total of 4.5k question-ranking pairs, of which 0.5k are unanswerable.\footnote{Examples of data samples with annotated information nuggets and answerability scores on various levels are provided in the repository accompanying the paper.}

Overall, our CAsT-answerability dataset contains binary answerability labels on three levels: sentence, passage, and ranking. 
Sentence- and passage-level answerability is divided into train (90\%), and test (10\%) portions; the splitting is done on the question level to avoid information leakage. Ranking-level answerability has only a test set.
See Table~\ref{tab:dataset_stats} for a summary.
\section{Answerability Detection}
\label{sec:unaswerability_detection}

\begin{figure}[t]
    \centering
    \includegraphics[width=1.\textwidth]{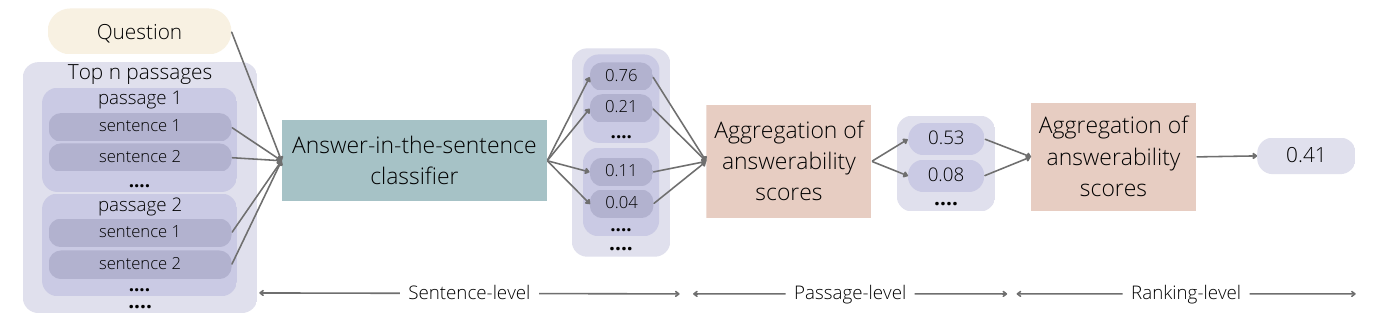}
    \caption{Overview of our answerability detection approach.}
    \label{fig:system_architecture}
\end{figure}

The challenge of answerability in CIS arises from the fact that the answer is typically not confined to a single entity or text snippet, but rather spans across multiple sentences or even multiple passages. 
Note that answerability extends beyond the general notion of relevance and asks for the presence of a specific answer. 
At the core of our approach is a sentence-level classifier that can distinguish sentences that contribute to the answer from ones that do not.  These sentence-level estimates are then aggregated on the passage level and then further on the ranking level (i.e., set of top-n passages) to determine whether the question is answerable; see~Figure~\ref{fig:system_architecture}.
Operating on the sentence level is a design decision that has the added benefit that a future summary generation step may take a filtered set of sentences that contribute to the final answer as input.

\subsection{Answer-in-the-Sentence Classifier}
\vspace*{-0.5\baselineskip}

The answer-in-the-sentence classifier is trained on sentence-level data from the train portion of the CAsT-answerability dataset. In some of the experiments, this data is augmented by data from the SQuAD 2.0 dataset~\citep{Rajpurkar:2018:ACLa} to provide the classifier with additional training material and thus guidance in terms of questions that can be answered with a short snippet contained in a single sentence. Data from SQuAD 2.0 is downsampled to be balanced in terms of the number of answerable and unanswerable question-sentence pairs. The classifier is built using a BERT transformer model with a sequence classification head on top (BertForSequenceClassification provided by HuggingFace\footnote{\url{https://huggingface.co/docs/transformers/model_doc/bert\#transformers.BertForSequenceClassification}}). Each data sample contains \texttt{question [SEP] sentence} as input and a binary answerability label. The output of the classifier is the probability that the sentence contains (part of) the answer to the question.

\subsection{Aggregation of Sentence-level Answerability Scores}
\vspace*{-0.5\baselineskip}

In reality, answers are not confined to a single sentence but can be spread across several passages. We thus need a method to aggregate results obtained from the sentence-level classifier to decide whether the question can be answered given (1) a particular passage or (2) a set of top-ranked passages, referred to as a ranking.

We consider two simple aggregation functions, \emph{max} and \emph{mean}, noting that more advanced score- and/or content-based fusion techniques could also be applied in the future~\citep{Kurland:2018:SIGIR}.
Intuitively, \emph{max} is expected to work particularly well for factoid questions where the answer is relatively short and usually contained in a single sentence, while \emph{mean} should capture the cases where pieces of the answer are spread over several sentences within the passage or across passages.
The aggregated answerability score for a given passage is compared against a fixed threshold; passages with an aggregated score exceeding this threshold are identified as containing the answer.
We set the threshold values on a validation partition (10\% of the dataset, sampled from the training partition); %.~\footnote{\tocheck{The CAsT-answerability dataset is divided into training, validation, and test partitions, encompassing 80\%, 10\%, and 10\% of the dataset respectively. The validation partition is sampled from the training partition to maintain the specified proportions, accounting for approximately 11\% of the training partition.}} of the CAsT-answerability dataset 
0.5 for max and 0.25 for mean. 

An analogous procedure is repeated for the top $n=3$ passages in the ranking to decide on ranking-level answerability. Here, the aggregation methods take the passage-level answerability scores as input (obtained using max or mean aggregation of sentence-level probabilities). The resulting values are compared against a fixed threshold (using the same values as for passage-level aggregation) to yield a final ranking-level answerability prediction. 
\section{Results}
\label{sec:results}
\vspace*{-0.25\baselineskip}

Table~\ref{tab:evaluation} presents the answerability results on the sentence-, passage-, and ranking-levels on the test partition of CAsT-answerability in terms of accuracy.

\begin{table}[t]
\small
\centering
\setlength\tabcolsep{3pt}
\caption{Answerability detection results in terms of classification accuracy. The best scores for each level are in boldface. For the augmented classifier (rows 5--8), significant differences against the respective method (rows 1--4) are indicated by $^*$. ChatGPT results are tested against the best classifier in rows 1--8. We use McNemar's test with $p<0.05$.} 
\label{tab:evaluation}
\begin{tabular}{l|c|c|c|c|c}
    \multirow{2}{*}{\textbf{Classifier}} & \textbf{Sentence} & \multicolumn{2}{c|}{\textbf{Passage}} & \multicolumn{2}{c}{\textbf{Ranking}} \\
    \cline{2-6}
     & \textbf{Acc.} & \textbf{Aggr.} & \textbf{Acc.} & \textbf{Aggr.} & \textbf{Acc.} \\  
    \hline
      \multirow{4}{*}{CAsT-answerability} & \multirow{4}{*}{0.752} & \multirow{2}{*}{Max} & \multirow{2}{*}{0.634} & Max & 0.790 \\
      \cline{5-6}
      &  &  &  & Mean & \textbf{0.891} \\
      \cline{3-6}
      &  & \multirow{2}{*}{Mean} & \multirow{2}{*}{0.589} & Max & 0.332 \\
      \cline{5-6}
      &  &  &  & Mean & 0.829 \\
      \hline
      \multirow{4}{*}{\begin{tabular} {@{}l@{}}CAsT-answerability\\\ augmented with \\\ SQuAD 2.0\end{tabular}} & \multirow{4}{*}{\textbf{0.779$^*$}} & \multirow{2}{*}{Max} & \multirow{2}{*}{0.676$^*$} & Max & 0.810$^*$ \\
      \cline{5-6}
      &  &  &  & Mean & 0.848$^*$ \\
      \cline{3-6}
      &  & \multirow{2}{*}{Mean} & \multirow{2}{*}{0.639$^*$} & Max & 0.468$^*$ \\
      \cline{5-6}
      &  &  &  & Mean & 0.672$^*$ \\
      \hline
      \multicolumn{3}{l|}{\multirow{2}{*}{ChatGPT passage-level (zero-shot)}} & \multirow{2}{*}{\textbf{0.787$^*$}} & T=0.33 & 0.839$^*$ \\
      \multicolumn{3}{l|}{} &  & T=0.66 & 0.623$^*$ \\
      \hline
      \multicolumn{5}{l|}{ChatGPT ranking-level (zero-shot)} & 0.669$^*$ \\
      \hline
      \multicolumn{5}{l|}{ChatGPT ranking-level (two-shot)} & 0.601$^*$ \\
    \end{tabular}
\end{table}

\emph{Does data augmentation help answerability detection?}
On the sentence level, we find that augmenting the CAsT-answerability dataset with additional training examples from SQuAD 2.0 improves performance. These improvements also carry over to the first aggregation step on the passage level. However, the best ranking-level results are obtained by aggregating results obtained from the classifier trained only on CAsT-answerability. 
It may result from the fact that  SQuAD 2.0 training data focuses on questions with short-span answers (like entities or numbers) confined to a single sentence. This could mislead the classifier to overlook answers spanning multiple sentences or passages. Thus, while sentence-level answerability prediction benefits from augmented data, this does not translate to effective passage or ranking-level answerability prediction.

\emph{Which of the two aggregation methods performs better?}
In all cases, max aggregation on the passage level followed by mean aggregation on the ranking level gives the best results. Intuitively, this configuration captures single sentences with high answerability scores in individual passages (max aggregation on passage level) that give a high average score for the whole ranking (mean aggregation on ranking level) for answerable questions.

\emph{How competitive are these baselines in absolute terms?}
Ours is a novel task, with no established baselines to compare against.
However, using a large language model (LLM) for generating the final response from the top retrieved passages is a natural choice. Therefore, for reference, we compare against a state-of-the-art LLM, using the most recent snapshot of GPT-3.5 (gpt-3.5-turbo-0301) via the ChatGPT API. 
We consider two settings: given a passage (analogous to the passage-level setup) and given a set of passages as input (analogous to the ranking-level setup). 
We prompt the model to verify whether the question is answerable in the provided passage(s) and return 0 or 1 accordingly.\footnote{The prompts are available in the repository accompanying the paper.}
In the passage-level setup, the passage-level predictions returned by ChatGPT are aggregated using fixed thresholds to obtain a ranking-level prediction. 
The max aggregation boils down to checking whether any of the passages is predicted to contain the answer. In the case of mean aggregation, a threshold of 0.33 or 0.66 (based on the fact that binary values are returned for passage-level answerability predictions) would mean that 1 or 2 out of 3 passages, respectively, contain the answer.
In the ranking-level setup, we experiment with both a zero-shot setting, where neither examples nor context is given to the model,
and a two-shot setting containing a question followed by two sentences (one positive and one negative example) extracted from the passage.
We observe that the passage-level answerability scores of ChatGPT are higher than ours, but after ranking-level aggregation, it is no longer the case. 
Further, performing the ranking-level task directly results in significantly lower performance.
These results indicate that LLMs have a limited ability to detect answerability without additional guidance. Our baseline methods trained on small datasets and based on simple classifiers with multi-step results aggregation turn out to be more effective for answerability prediction and thus represent a strong baseline.
\section{Conclusion}
\label{sec:concl}

Unanswerable questions pose a challenge in conversational information seeking. 
To study this problem, we have developed a test collection, based on two editions of the TREC CAsT benchmark, with sentence-, passage-, and ranking-level answerability labels.
We have also presented a baseline approach based on the idea of sentence-level answerability classification and multi-step results aggregation, and evaluated multiple instantiations of this approach with different configurations. 
Despite their simplicity, our baselines have been shown to outperform a state-of-the-art LLM on the task of answerability prediction.

In this paper, we have simplified the scenario by treating answerability as binary concept: a question is answerable if any sentence in the returned passages contains the answer. In practice, answerability is more nuanced, with some pieces of the information found but not all. A more realistic future approach would involve an ordinal scale (e.g., unanswerable, partially answerable, fully answerable), which would necessitate ground truth assessments with an explicit specification of the different facets/aspects of the answer. We are not aware of any dataset for information-seeking tasks (conversational or not) that would provide this information.

\subsubsection{Acknowledgments}

This research was supported by the Norwegian Research Center for AI Innovation, NorwAI (Research Council of Norway, project number 309834).

\renewcommand*{\bibfont}{\scriptsize}
\bibliographystyle{splncs04}
\bibliography{ecir2024-unanswerability.bib}

\end{document}